\documentclass[oneside,a4paper]{article}

\usepackage{graphicx} 
\graphicspath{ {./images/} }

\usepackage{hhline}

\usepackage{comment}

\usepackage{hyperref}
\usepackage{amssymb}
\usepackage{authblk}
\providecommand{\keywords}[1]{\textbf{\textit{Keywords:}} #1}

\title{\textbf{Comparing Flexible Skylines And Top-k Queries: Which Is the Best Alternative?}}
\author{Flavio, FR, Rizzoglio}
\affil{Politecnico di Milano\\
Milan, Italy\\
\href{mailto:flavio.rizzoglio@mail.polimi.it}{flavio.rizzoglio@mail.polimi.it}}
\date{$8^{th}$ February 2022}

\begin{document}
\maketitle
\begin{abstract}
The question of how to get the best results out of the data we have is an everlasting problem in data science. The two main approaches to tackle the problem are top-k queries and skyline queries. Since their introduction, a new paradigm called flexible skylines has emerged. The aim of this survey is to provide a solid comparison between the new and the old approaches, understanding and exploring their differences and similarities.
\end{abstract}


\section{Introduction}
Top-k queries are the first solution that has been proposed to solve the problem of finding the best results in data. In fact, in different applications, users may want to access or visualize only the most relevant results from an otherwise large set of tuples. Top-k queries aim to identify the top results through the use of a scoring function, defined by the preferences of the user. The better the score, the higher an object is in the output data. Top-k queries could be classified based on different dimensions such as Query Model, Data Access Methods, Implementation Level, Data and Query Uncertainty and Ranking Function \cite{ilyas2008survey}.
\newline
Sometimes, however, it is not possible to compute the best k results. For example, let’s assume that we want to book a train between Milan and Rome and we are looking for a cheap ticket on a high-speed train. As we might expect, high-speed trains tend to be more expensive than normal ones, so we cannot properly classify the tickets as cheap. The purpose of skyline queries \cite{skylineoperator} is to solve this problem by displaying the most interesting results to the end-user. Skylines are based on the concept of dominance, meaning that a point dominates another point only if it is as good or better in all the dimensions and better in at least one dimension. Points that aren’t dominated by other points make up the skyline.
\newline
Finally, flexible skylines or F-skylines \cite{ciaccia2017reconciling} (in the original paper called R-skylines) are an evolution of traditional skyline queries. F-skylines implement the concept of dominance as normal skylines but apply it to a set of scoring functions, thus allowing greater flexibility than the techniques previously described. 
\newline
To fully understand flexible skylines, we must first introduce the concept of \emph{F}-dominance. A generic tuple \emph{s} \emph{F}-dominates another tuple \emph{z} if, for every scoring function taken into consideration, the tuple \emph{s} is no worse than \emph{z}. Giving a more precise definition:
\paragraph{}\emph{Def 1.1:} Let $F$ be a set of monotone scoring functions. A generic tuple $s$ \emph{F}-dominates another tuple $s \neq  t$, indicated by $s \prec_F z$, if $\forall f \in F$ . $f(s) \leq f(z)$ (bearing in mind that if $f(s) \leq f(z)$, $f(s)$ has a better or equal result than $f(z)$).
\newline
\newline
Flexible skylines use two operators, both subsets of the skyline: ND, the set of non-\emph{F}-dominated tuples and PO, the set of potentially optimal tuples that returns the best tuples (i.e., top-1) according to some scoring function in $F$. It is important to note that PO is always a subset of ND.
\newline
Other notable approaches that try to evolve the traditional paradigms and bridge top-k and skylines are ORD/ORU \cite{mouratidis2021marrying} and UTK (Uncertain Top-k Query) \cite{mouratidis2018exact}. ORD implements the OSS (output-size specified) property, relaxed input and it is dominance-oriented, while the ORU approach also focuses on the first two characteristics of ORD as well, but it follows a utility paradigm instead of a dominance-oriented paradigm.
\newline
UTK has two versions. In the first one $UTK_{1}$, given a set of uncertain user preferences that form an approximate description of the weight values, it reports all options that may belong to the top-k set, while in the second $UTK_{2}$ it additionally includes in the result set the exact top-k set for each of the possible weight settings.
\newline
\newline
In the next sections, we will discuss how these techniques can be compared, their main strengths and weaknesses and some of their applications.

\section{Comparisons}
As argued in Reference \cite{freitas2004critical}, each approach has its own strengths and weaknesses, which can be summarised in \emph{Table~\ref{table:1}}. Top-k queries, since they are so tightly bound to user preferences, lack the simplicity of formulation and the ability to display interesting results.
\newline
For example, the best plane ticket is determined by the scoring function \emph{P = 0.9*Price + 0.1*NumOfStopovers}, the function P could yield good results for a person whose main focus keeping the price as low as possible. However, a businessman that flies a lot might object and, to save time, may want to prioritize diminishing as low as possible the number of stopovers. This, of course, would result in a modification of the weights, yielding a completely different result.
\newline
As much as this structure can hinder simplicity and the display of overall interesting results, it has also a lot of benefits. Top-k queries display an exceptional level of control of the cardinality of the results while offering trade-off among attributes and efficiency.
\newline
In skylines, there is no need of specifying the weights of a scoring function, thus allowing for a more straightforward and simple to use paradigm. This technique also focuses greatly on the effort of providing interesting results, a task that is not achievable with the use of traditional ranking queries. However, skylines present some weaknesses too: the result of a skyline operation may contain a large number of tuples and they can’t provide any form of user preferences.
\newline
Overall, skylines trade user preferences and efficiency for the simplicity of use and the ability to show interesting results.
\newline
Flexible skylines, combining the two techniques, try to bridge the two previous approaches \cite{ciaccia2020flexible}, resulting in a more balanced approach. F-skylines can, in fact, account for the different importance of various attributes. F-skylines can also consider and model user preferences by leveraging the weights used in different scoring functions, while still displaying the overall most interesting results.
\newline
We also find that the ORD/ORU approach has a lot of benefits compared to traditional techniques, mainly: personalization, controllable output size and preference specification flexibility. These techniques are not as simple as F-skylines but, on the other hand offer a better performance.
\newline
On the topic of alternative approaches UTK focuses on providing the user with a practical and easy to use design by computing weight regions to take into account the uncertain preferences that a user may have. Basically, with this approach, we forego the control on the result cardinality in order to offer a better user experience.
\newline
\newline
We will now focus on one last dimension of analysis considered in \emph{Table~\ref{table:1}}: performance. As presented in the introductory section, flexible skylines are an evolution of traditional skylines. Therefore, we will only compare the performance of top-k queries and F-skylines.

\begin{table}
\centering
\begin{tabular}{|l||l|l|l|l|l|} 
\hline
\textbf{Dimension of analysis}          & \textbf{Top-k} & \textbf{Skyline} & \textbf{F-skylines}                    & \textbf{ORD/ORU }      & \textbf{UTK}       \\ 
\hhline{|=::=====|}
Control of result cardinality  & Yes   & No      & By modifying constraints & Yes           & No        \\ 
\hline
Trade-off among attributes     & Yes   & No      & Yes                           & Yes           & Yes       \\ 
\hline
Simplicity                     & No    & Yes     & Yes                           & No            & Yes       \\ 
\hline
Display of interesting results & No    & Yes     & Yes                           & Yes           & Yes       \\ 
\hline
Efficient performance          & Yes   & No      & No        & No  & No  \\
\hline
\end{tabular}
\caption{Strengths and Weaknesses of the techniques presented.}
\label{table:1}
\end{table}

\subsection{Performance of Top-k queries}
\label{sec:sec2.1}
Firstly, we introduce the notion of instance optimality. Instance optimality is a form of optimality aimed at when standard optimality is unachievable.
\paragraph{}\emph{Def 2.1.1:} Let $A$ be a family of algorithms, $I$ a set of problem instances and $cost$ be a cost metric applied to an algorithm-instance pair. Algorithm $A^{*}$ is instance-optimal wrt $A$ and $I$ for the cost metric $cost$ if there exist constants $k_{1}$ and $k_{2}$ such that, for all $A_{1}\in A$ and $I_{1}\in I$, such that: $cost(A^{*}, I_{1}) \leq k_{1} \cdot cost(A_{1}, I_{1}) + k_{2}$
\newline
\newline
Top-k queries are known for their high efficiency. The most established and popular algorithms that allow ranking queries to reach this goal are: Fagin’s Algorithm \cite{fagin1998fa} (or FA), Threshold Algorithm \cite{FAGIN2003614} (or TA) and No Random Access \cite{FAGIN2003614} (or NRA), with the latter one only used when random access cannot be executed since its performance is worse than FA and TA. 
\newline
FA is a non-instance optimal algorithm, has a sub-linear time complexity and its stopping criterion is independent of the scoring function. FA has three phases: In the first phase, the algorithm only executes sorted accesses, while in the second phase only random accesses are processed. Then, in the third and final phase, the scores of the retrieved objects are computed using the chosen scoring function finally determining the top-k results. FA is greatly surpassed by the instance optimal algorithm TA, which in general, performs much better since it can adjust to a specific scoring function. The TA algorithm is instance optimal and utilizes a threshold value $f(\tau)$ calculated by applying the chosen scoring function $f$ to the tuple $\tau$ composed of the last scores seen by sorted access on each ranking. Also, differently from FA, random accesses are performed after executing the sorted accesses (one for each ranking) and not in a second phase.
\newline
Since their first introduction, however, many algorithms that improve the efficiency of these baseline algorithms have been published. BPA (Best Position Algorithm) and its improved version BPA2 \cite{akbarinia2007best} provide such examples. As explained in Reference \cite{akbarinia2007best}, with BPA2 we can outperform TA by a factor of $\frac{(m+1)}{2}$, where $m$ is the number of the $m$ best positions. As evident, BPA2 becomes more efficient the larger the required output size is. These kinds of problems are not uncommon in real applications of top-k queries. For example, let’s consider a database that holds every branch of an international bank and we want to view the subsidiaries with the most transactions made. Since the database is very large, the number of top positions may range from a few tens to the order of thousands.
\newline
It is important to note that FA, TA, NRA and BPA2 are algorithms designed for a distributed setting where multiple components of the systems are located on different networked computers. On \emph{Table~\ref{table:2}} we find a summary of all the main top-k algorithms discussed, ranked by efficiency. 

\begin{table}
\centering
\begin{tabular}{|l||l|l|} 
\hline
\textbf{Algorithms} & \textbf{Data Access}              & \textbf{Notes}                                                           \\ 
\hhline{|=::==|}
BPA2       & Sorted and random access & Better than TA by a factor of (m+1)/2                           \\ 
\hline
TA         & Sorted and random access & Instance-optimal, can adjust to a specific scoring function  \\ 
\hline
FA         & Sorted and random access & Stopping criterion is independent from the scoring function     \\ 
\hline
NRA        & Sorted access            & Instance-optimal but no exact scores                            \\
\hline
\end{tabular}
\caption{Summary of main top-k algorithms, ranked by efficiency (best BPA2).}
\label{table:2}
\end{table}

\subsection{Performance of Flexible Skylines}
Before delving into F-Skylines, let’s discuss two of the most famous algorithms for traditional skylines: Block Nested Loop (BNL) \cite{skylineoperator} and Sort-Filter-Skyline (SFS) \cite{chomicki2003skyline}, both designed for centralized settings. BNL was the first algorithm introduced to compute skylines. It adopts a naive approach, simply applying a nested-loops algorithm and confronting each tuple with all the other tuples. This, of course, results in a very inefficient algorithm, particularly for large datasets. Overall, BNL has a time complexity of $O(N^{2})$. A solution to this problem is SFS, which implements pre-sorting to improve efficiency in big datasets. Even though SFS is much more efficient in some cases, the overall complexity is still the same as BNL.
\newline
\newline
Now, we will introduce some of the algorithms initially proposed for F-skylines: SVE1F and PODI2 and then compare them to SFS.
\newline
SVE1F is an algorithm used for computing ND. It is a one-phased algorithm, meaning that ND is directly calculated from the database and the skyline is not computed beforehand. SVE1F also implements a pre-sorting of the dataset before any operation is carried out and uses vertex enumeration to perform dominance tests. SVE1F has a worst-case time complexity of $O(ve(c) + N(logN + |ND| * q))$, where $O(ve(c))$ is the time complexity of the vertex enumeration, $|ND|$ is the number of tuples in the final ND set and q is the number of vertices of used in the vertex enumeration.
\newline
PODI2 is an algorithm used for computing PO. It is a two-phased algorithm that, in the first phase, calculates the ND tuples and then, in the second phase, filters out all the non-PO tuples, finally outputting the PO set. PODI2 also utilizes an incremental approach to test whether a tuple is potentially optimal or not. This test requires solving a linear programming (LP) problem, which can be potentially highly inefficient and time consuming when the dataset is very large. To solve this problem an incremental approach is used, which implies solving the complete LP problem by solving smaller LP problems of increasing size. The worst-case time complexity is $C_{nd} + O(|ND| * log|ND| * lp(q, |ND|))$, where $C_{nd}$ is the time complexity of computing ND and $O(lp(q, |ND|))$ is the time complexity of the F-dominance tests. Both SVE1F and PODI2 are designed for centralized systems.
\newline
In Reference \cite{ciaccia2020flexible}, a comparison between SFS, SVE1F and PODI2 is carried out by testing the algorithms on three different datasets:
\begin{itemize}
    \item \textbf{NBA}: a famous dataset with thousands of points which include different statistics on NBA players’ performance;
    \item \textbf{ANT}: a synthetic dataset with anticorrelated values across different dimensions;
    \item \textbf{UNI}: a synthetic dataset with uniformly distributed values.
\end{itemize}
In \emph{Table~\ref{table:3}}, we find a summary of this performance comparison, including the complexities of each algorithm. Also, in \emph{Figure 1}, the performance results are graphically displayed.
\begin{table}
\centering
\begin{tabular}{|l||l|l|l||l|} 
\hline
\textbf{Algorithm} & \textbf{NBA} & \textbf{UNI} & \textbf{ANT} & \textbf{Complexity } \\ 
\hline
SFS                & 0,58         & 0,34         & 11,17        & $O(n^{2})$ \\
SVE1F              & 0,65         & 0,36         & 1,54         & $O(ve(c) + N(logN + |ND| * q))$ \\
PODI2              & 1,74         & 1,41         & 12,69        & $C_{nd} + O(|ND| * log|ND| * lp(q, |ND|))$ \\
\hline
\end{tabular}
\caption{Summary of the comparison between SFS, SVE1F and PODI2, values are in seconds.}
\label{table:3}
\end{table}
\begin{figure}[ht]
    \centering
    \includegraphics[scale=0.9]{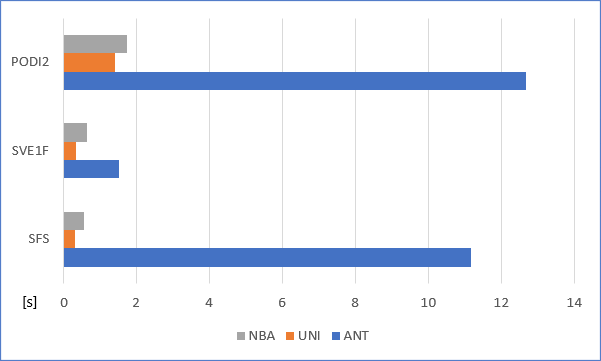}
    \caption{Performance comparison between SFS, SVE1F, PODI2}
\end{figure}
\newline
As we can see, calculating ND and PO can be a potentially detrimental task that could hinder performance. All the algorithms initially proposed to calculate these two operators had a higher time complexity than top-k algorithms, resulting in less efficiency. This is of course a big issue when comparing F-skyline with top-k queries as this could render the added benefits of F-skylines not worth the additional computing time. 
\newline
Moving again to a distributed setting, some progress has been made with the introduction of the Flexible Score Aggregation (FSA) \cite{ciaccia2018fsa}, algorithm. FSA is an instance optimal algorithm and it is the result of a combination between the FA and TA algorithms introduced in \hyperref[sec:sec2.1]{\emph{Section 2.1}}. FSA allows the user to calculate the set $ND_{k}$ which, when $k$ is equal to one, matches the result of ND. This algorithm is very efficient and effective, leading to a reduction in the result set within cardinality, even for large datasets. However, this is not enough, since FSA still presents a quadratic complexity, while top-k queries are far better in terms of raw performance.
\newline
\newline
Finally, comparing the performance of F-skylines with top-k queries we can see that F-skylines are not comparable with the level of performance of top-k queries. Bearing in mind that flexible skylines and top-k algorithms are designed for different setting, it is clear that the main objective of these new approaches is not maximizing performance. For example, calculating PO is still really inefficient, albeit offering much more interesting results than any other traditional technique. The topic of the quality and content of the output data will be covered in \hyperref[sec:sec2.4]{\emph{Section 2.4}}. 
\newline
\newline
Regarding performance, a notable observation is that dominance tests efficiency could be further improved using decision trees \cite{choi2021optimization}. However, this methodology has not yet been tested with F-skylines, so it remains an open problem.

\subsection{Comparisons between the new approaches}
Shifting the focus from the comparison with traditional top-k, we now compare F-skylines with other two new approaches, ORD/ORU and UTK. 
\newline
The main difference between ORD/ORU and the approaches at \cite{ciaccia2020flexible} and \cite{mouratidis2018exact} is that ORD/ORU produce an output of controllable size, respecting the OSS property, as seen in \emph{Table~\ref{table:1}}. Also, F-skylines and UTK utilize a fixed polytope region given in advance to establish dominance, whereas the ORD/ORU dominance region is effectively a hyper-sphere and usually not given in advance. Therefore, as established in \cite{mouratidis2021marrying}, ORD/ORU are hardly comparable, in terms of output data, to the other two techniques. However, focusing on performance only, ORD/ORU algorithms are much more efficient than F-skylines and UTK. ORD, in particular, is 2 to 4 orders of magnitude faster than F-skylines while ORU is 12 to 134 (or even more in some cases) faster than JAA, the best algorithm for fixed region approaches, employed by UTK.
\newline
Another important comparison is that F-skylines, with its operators ND and PO, generalize top-k queries fixing $k = 1$, while UTK and the ORD/ORU approaches allow any $k$.

\subsection{Output data}
\label{sec:sec2.4}
In order to discuss differences in output data, we first have to introduce two measures: \emph{precision} and \emph{recall}. Let $TOP_{K}$  be a set of top-k tuples on a dataset $D$ and $S$ be the set of the tuples present in $SKY$, $ND$ and $PO$, over the same dataset $D$, where $SKY$ is the skyline output set on $D$. We define precision as $PRE(S) = \frac{|S \cap TOP_{K}|}{k}$ and recall as $REC(S) = \frac{|S \cap TOP_{K}|}{|S|}$. 
\newline
In the context of this comparison, the aim is to assess whether or not the results yielded by the two techniques overlap in a significant way or if they present different outputs. The value of this two measures is highly dependent on how similar the top-k scoring function is to the family of scoring functions used in the F-skyline. In \emph{Table~\ref{table:4}}, we can see what precision and recall values represent in our context.
\newline
Regarding the precision, experimental data gathered in \cite{ciaccia2020flexible} shows that, for each k in a generic dataset $D$, $PRE(PO) \leq PRE(ND) \leq PRE(SKY)$, meaning that a result set of tuples of a top-k, will more likely have SKY tuples than ND or PO tuples. This means that the more we seek interesting results with F-skylines, the more the precision will drop when comparing to a a traditional top-k approach.
\newline
Regarding the recall measure, experimental data of Reference \cite{ciaccia2020flexible} finds that, when the number of tuples from the top-k and F-skylines operators is equal, the range of $REC(ND)$ is between 38\% and 50\%, meanwhile for $REC(PO)$ the range drastically decreases to 18\%–38\%. This means that, in order for top-k to reach a quality of output level similar to that of F-skylines, we would need an extremely large k, resulting in a huge output set.

\begin{table}
\centering
\begin{tabular}{|l||l|l|} 
\hline
\textbf{Measure}   & \textbf{High Value}                                    & \textbf{Low Value}                                                \\ 
\hline
Precision & Most top-k tuples are also in S        & Most top-k tuples are \underline{NOT} also in S         \\ 
\hline
Recall    & Most tuples in S are also top-k tuples & Most tuples in S are \underline{NOT} also top-k tuples  \\
\hline
\end{tabular}
\caption{Summary of PRE and REC values.}
\label{table:4}
\end{table}

\section{Applications}
In this section, we will present some of the main applications of both top-k queries and F-skylines. Regarding top-k queries, they are used in a wide variety of fields, thanks to their efficiency and relatively straightforward approach.
Two of the most classical applications for top-k are: multi-criteria queries and k-nearest neighbours. An example of multi-criteria queries could be ranking the best k cars in an online car dealer, combining different criteria like mileage, price, color etc. Instead, k-nearest neighbours is a similarity search used in machine learning and it is utilized to, given N points in a D-dimensional space and a point Z in that space, find the k points closest to point Z.
\newline
Additionally to these classical applications, top-k queries are used in a multitude of other niche applications, for example, privacy-preserving top-k queries \cite{vaidya2005privacy} or reverse top-k queries \cite{vlachou2010reverse}, where the first technique aims to bridge privacy with top-k while the second technique tries to look at top-k queries through the lenses of the product manufacturer and not the user.
\newline
F-skylines can, of course, be used in all the fields where traditional skylines are useful and established, bearing in mind that the number of real applications is smaller, due to the most recent introduction of the concept. However, a very important field where F-skylines could have a huge impact are Data Warehouses, where queries already take up a lot of time and where finding useful and interesting data is extremely useful. Data Warehouses are often centralized and collect a huge amount of data from different sources, resulting in really expensive queries. However, this is accepted and common, since the main objective of Data Warehouses is not performance, but extracting meaningful and interesting data to be used in a business context, aligning particularly well with the purpose and goals of Flexible skylines.

\section{Conclusions}
In this paper, we tried to compare the ranking techniques with the hope of finding out whether top-k is better than flexible skylines, or vice versa.
\newline
There is no definitive answer, as it greatly depends on the dimension of evaluation. In general, we conclude that top-k is better if the main focus is on pure performance and computing capabilities. If we instead focus on usability, F-skylines offer the user a more simple approach, without leaving behind user preferences. On the topic of output data, F-skylines are clearly the better alternative, giving the most interesting results in a more compact result set.
\newline
\newline
Overall, F-skylines are better in every aspect apart from performance, basically trading efficiency for greater simplicity and better output data. Notably, also ORD/ORU and UTK are better than top-k in every aspect apart from performance.

\bibliographystyle{plain}
\bibliography{refs.bib}

\begin{thebibliography}{10}

\bibitem{akbarinia2007best}
Reza Akbarinia, Esther Pacitti, and Patrick Valduriez.
\newblock Best position algorithms for top-k queries.
\newblock In {\em international conference on Very large data bases (VLDB)},
  pages 495--506. ACM, 2007.

\bibitem{skylineoperator}
S.~Börzsönyi, D.~Kossmann, and K.~Stocker.
\newblock The skyline operator.
\newblock {\em ICDE}, pages 421--430, 2001.

\bibitem{choi2021optimization}
Jong-Hyeok Choi, Fei Hao, Yoo-Sung Kim, and Aziz Nasridinov.
\newblock Optimization of dominance testing in skyline queries using decision
  trees.
\newblock {\em IEEE Access}, 9:130170--130184, 2021.

\bibitem{chomicki2003skyline}
Jan Chomicki, Parke Godfrey, Jarek Gryz, and Dongming Liang.
\newblock Skyline with presorting.
\newblock In {\em ICDE}, volume~3, pages 717--719, 2003.

\bibitem{ciaccia2017reconciling}
Paolo Ciaccia and Davide Martinenghi.
\newblock Reconciling skyline and ranking queries.
\newblock {\em Proceedings of the VLDB Endowment}, 10(11):1454--1465, 2017.

\bibitem{ciaccia2018fsa}
Paolo Ciaccia and Davide Martinenghi.
\newblock {FA} + {TA} $<$ {FSA}: {F}lexible score aggregation.
\newblock In {\em Proceedings of the 27th ACM International Conference on
  Information and Knowledge Management}, pages 57--66, 2018.

\bibitem{ciaccia2020flexible}
Paolo Ciaccia and Davide Martinenghi.
\newblock Flexible skylines: Dominance for arbitrary sets of monotone
  functions.
\newblock {\em ACM Transactions on Database Systems (TODS)}, 45(4):1--45, 2020.

\bibitem{fagin1998fa}
Ronald Fagin.
\newblock Fuzzy queries in multimedia database systems.
\newblock In {\em Proceedings of the Seventeenth ACM SIGACT-SIGMOD-SIGART
  Symposium on Principles of Database Systems}, PODS '98, page 1–10, New
  York, NY, USA, 1998. Association for Computing Machinery.

\bibitem{FAGIN2003614}
Ronald Fagin, Amnon Lotem, and Moni Naor.
\newblock Optimal aggregation algorithms for middleware.
\newblock {\em Journal of Computer and System Sciences}, 66(4):614--656, 2003.
\newblock Special Issue on PODS 2001.

\bibitem{freitas2004critical}
Alex~A. Freitas.
\newblock A critical review of multi-objective optimization in data mining: A
  position paper.
\newblock {\em SIGKDD Explor. Newsl.}, 6(2):77–86, dec 2004.

\bibitem{ilyas2008survey}
Ihab~F Ilyas, George Beskales, and Mohamed~A Soliman.
\newblock A survey of top-k query processing techniques in relational database
  systems.
\newblock {\em ACM Computing Surveys (CSUR)}, 40(4):1--58, 2008.

\bibitem{mouratidis2021marrying}
Kyriakos Mouratidis, Keming Li, and Bo~Tang.
\newblock Marrying top-k with skyline queries: Relaxing the preference input
  while producing output of controllable size.
\newblock In {\em Proceedings of the 2021 International Conference on
  Management of Data}, pages 1317--1330, 2021.

\bibitem{mouratidis2018exact}
Kyriakos Mouratidis and Bo~Tang.
\newblock Exact processing of uncertain top-k queries in multi-criteria
  settings.
\newblock {\em Proceedings of the VLDB Endowment}, 11(8):866--879, 2018.

\bibitem{vaidya2005privacy}
Jaideep Vaidya and Chris Clifton.
\newblock Privacy-preserving top-k queries.
\newblock In {\em 21st International Conference on Data Engineering (ICDE'05)},
  pages 545--546. IEEE, 2005.

\bibitem{vlachou2010reverse}
Akrivi Vlachou, Christos Doulkeridis, Yannis Kotidis, and Kjetil N{\o}rv{\aa}g.
\newblock Reverse top-k queries.
\newblock In {\em 2010 IEEE 26th International Conference on Data Engineering
  (ICDE 2010)}, pages 365--376. IEEE, 2010.

\end{thebibliography}
\end{document}